\documentclass[5p]{elsarticle}

\usepackage[dvipsnames]{xcolor}
\usepackage{hyperref}
\hypersetup{hidelinks}

\biboptions{sort&compress}
\begin{document}

\begin{frontmatter}

\title{
Semi-relativistic antisymmetrized molecular dynamics for energetic neutron production in intermediate energy heavy-ion reactions
}

\author[address1,address2]{Q. Hu
\corref{mycorrespondingauthor}}
\cortext[mycorrespondingauthor]{Corresponding authors}
 \ead{qianghu@impcas.ac.cn }

\author[address1]{G. Y. Tian}

\author[address3]{R. Wada
\corref{mycorrespondingauthor}}
 \ead{wada@comp.tamu.edu}

\author[address4]{X. Q. Liu} 
\author[address4]{W. P. Lin}

\author[address5]{H. Zheng}

\author[address1,address2]{Y. P. Zhang}
\author[address1]{Z. Q. Chen} 
\author[address1]{R. Han}

\author[address6]{M. R. Huang}

 \address[address1]{Institute of Modern Phyiscs, Chinese Academy of Sciences, Lanzhou 730000, China}
 \address[address2]{University of Chinese Academy of Sciences, Beijing 100049, China}
 
 \address[address3]{Cyclotron Institute, Texas A$\&$M University, College Station, Texas 77843, USA}
 
 \address[address4]{Key Laboratory of Radiation Physics and Technology of the Ministry of Education, Institute of Nuclear Science and Technology, Sichuan University, Chengdu 610064, China}

 \address[address5]{School of Physics and Information Technology, Shaanxi Normal University, Xi’an 710119, China}

 \address[address6]{College of Physics and Electronics information, Inner Mongolia University for Nationalities, Tongliao, 028000, China}

\begin{abstract}
Relativistic corrections have been made in the non-relativistic antisymmetrized molecular dynamics (AMD) simulations to apply to the high energy neutron production in the $^{12}$C+$^{12}$C and $^{16}$O+$^{12}$C collisions at incident energies of 290 and 400 MeV/nucleon. The corrections are made in kinematics alone and no nucleon-nucleon inelastic scatterings nor meson productions are taken into account, and AMD with the relativistic corrections is called semi-relativistic AMD. 
The three-nucleon collision (3NC) and Fermi boost in the collision processes are taken into account in the non-relativistic AMD.  
Since the relativistic corrections tend to compensate in each other, the difference between the semi-relativistic and non-relativistic results become small.  
High energy tails of the available experimental neutron double differential cross sections, especially at larger angles, are well reproduced by AMD with the 3NC term both with non-relativistic and semi-relativistic simulations. 
These results indicate that the high energy neutrons are dominantly produced by the 3NC process in this incident energy range.
\end{abstract}


\end{frontmatter}

%\linenumbers

\section*{I. Introduction}

Intermediate heavy-ion collisions in the energy range of tens MeV to a few GeV/nucleon, evolving rapidly from preequilibrium stages of compression and expansion to the deexcitation phase, allow us to investigate non-equilibrium dynamics of finite-size fermion systems as well as the nature of dense and hot nuclear systems~\cite{bertsch1988,cassing1990}. The production of subthreshold mesons, high energy photons and nucleons, which occurs mainly at the beginning of the reaction, can offer information on the nuclear dynamics at the preequilibrium stages~\cite{cassing1990,gelbke1999,sapienza2001}. One of them is the three nucleon interaction. 

The three nucleon interaction consists of two parts, an attractive part and a repulsive part~\cite{friedman1981}. The attractive part is typically expressed by two-pion exchange with excitation of an intermediate $\Delta$ resonance following the Fujita-Miyazawa diagram~\cite{fujita1957} and is important at normal and subnormal densities. 
The repulsive part of the three nucleon process becomes important in heavy-ion reactions at intermediate energies above 100 MeV/nucleon.
However, in most of transport models, only two body interaction and binary collision term have been implemented.  
An attempt was made by Bonasera {\it et al.} in Ref.~\cite{bonasera1994} to extend a Boltzman-Nordheim-Vlasov (BNV) to include the three nucleon process in a three-nucleon collision (3NC) term, since the three-body potential part can be included in the momentum-dependent effective interaction ~\cite{bohnet1988}. In the model, the 3N collisions can occur when three particles are found inside the same interaction sphere, which is given by the 3NC cross section. In Ref.~\cite{wada2017}, one of the present authors, followed this scenario, but used the AMD framework, and was able to reproduce rather well the high energy proton spectra from the BEVALAC experiments at the incident energy up to 137 MeV/nucleon. 

On the experimental side, the energy spectra of fast protons from Ar+Ta collisions at 94 MeV/nucleon were measured with the $4\pi$ BaF$_{2}$ detector array MEDEA~\cite{migneco1992} at GANIL, France in 1990's, stimulated by the experimental reports of the surprisingly large cross section for the sub-threshold kaon production in Refs.~\cite{julien1991,legrain1999,lecolley1995}. Although the extracted sub-threshold kaon production cross sections were far below the values reported earlier, the high energy proton production studies were extended to other reaction systems, using $^{36,40}$Ar, $^{58}$Ni, $^{132}$Xe  beams on $^{51}$V, $^{58}$Ni, $^{98}$Mo, Ta, Au targets ~\cite{germain1997,sapienza1998,sapienza2004,coniglion2000}. In these studies, it was found that the protons with the energy three to four times larger than the beam energy per nucleon, were observed over a broad angular range. The measured energetic proton spectra and angular distributions were extended well above the kinematic limit, but their results were unable to be reproduced with standard BNV calculations~\cite{germain1997}. However, when they added a 3NC contribution based on a perturbed method of the BNV calculations with a sharp cut-off of the Fermi momentum in the initial nuclei, they were able to reproduce the energetic protons reasonably well ~\cite{germain1997}.

For the higher incident energies, two experimental data sets are available for the high energy nucleon studies. The experiments were carried out for different beam species on different targets at 290 - 600 MeV/nucleon at the Heavy Ion Medical Accelerator (HIMAC) facility in the National Institute of Radiological Sciences (NIRS) in Japan, to provide the precise neutron production cross sections in the neutron energy range of 1 MeV to several hundreds MeV for cancer therapy ~\cite{iwata2001,satoh2011}. It is quite interesting to apply the above models to high energy neutron productions, since the density of the overlap zone between the projectile and target becomes higher at an early stage of collisions, which enhances the 3NC process, and pion production may start to have impact on the dynamical process. 
However the dynamics in the available AMD code is performed in non-relativistic form and the relativistic treatment may become crucial at the incident energy above 100 MeV/nucleon. 
We treat the relativistic effects as the corrections to the non-relativistic AMD calculations for $^{12}$C + $^{12}$C and $^{16}$O + $^{12}$C at 290 and 400 MeV/nucleon to study the high energy neutron production. The results of AMD with the relativistic corrections as well as the non-relativistic AMD are compared with the experimental data.
This paper is organized as follows.
The modified AMD models are briefly described in section II. Detail comparisons of high energy neutron spectra and angular distributions are carried out in section III. A summary is given in section IV.

\section*{II. Modified antisymmetrized molecular dynamics model}

\subsection*{II-1. AMD}

%\textcolor{blue}{
In the AMD model, the wave function for an $A$-nucleon system is described by a Slater determinant $|\Phi\rangle$,
\begin{equation} \label{eq1}
|\Phi\rangle = \frac{1}{\sqrt{A!}}\det[\varphi_i(j)],
\end{equation}
where $\varphi_i=\phi_{Z_i}\chi_{a_i}$. The spin-isospin state $\chi_{a_i}$ of each single-particle state takes $p\uparrow, p\downarrow,n\uparrow$, and $n\downarrow$.
The spatial wave functions of nucleons $\phi_{Z_i}$ are given by a Gaussian wave function,
\begin{equation}\label{Eq2}
\langle{\bf r|\phi_{Z_i}}\rangle = \left(\frac{2\nu}{\pi}\right)^{3/4}\exp\left[-\nu\left({\bf r}-\frac{{\bf Z}_i}{\sqrt{\nu}}\right)^2+\frac{1}{2}{\bf Z}^2_i\right],
\end{equation}
where the width parameter $\nu=0.16$ fm$^{-2}$~\cite{ono1992} is a constant parameter common to all the wave packets. Thus the complex variables
$Z\equiv\{{\bf Z}_i;i=1,...A\}=\{{Z}_{i\sigma};i=1,...A,\sigma=x,y,z\}$ represent the centroids of the wave packets.
%Up to the antisymmetrization effect, the real part and the imaginary part of ${\bf Z}_i$ correspond to the centroids of the position and the momentum, respectively,
%\begin{equation}
%{\bf Z}_i = \sqrt{\nu}{\bf D}_i+\frac{i}{2\hbar\sqrt{\nu}}{\bf K}_i,
%\end{equation}
%where {\bf D} = $\langle\phi_Z|{\bf r}|\phi_Z\rangle/\langle\phi_Z|\phi_Z\rangle, {\bf K} = \langle\phi_Z|{\bf p}|\phi_Z\rangle/\langle\phi_Z|\phi_Z\rangle.$
%The AMD wave function $|\Phi\rangle$ contains many quantum features in it and well describes the ground states of nuclei.
%}

%\textcolor{blue}{
The time evolution of the wave packet parameters $Z$ is determined by the time-dependent variational principle and the two-nucleon collision process.
The former is described as
\begin{equation}\label{Eq3}
\delta\int dt\frac{\langle\Phi(Z)|(i\hbar\frac{d}{dt}-H)|\Phi(Z)\rangle}{\langle\Phi(Z)|\Phi(Z)\rangle} = 0.
\end{equation}
The equation of motion for $Z$ derived from the time-dependent variational principle is
\begin{equation}\label{Eq4}
i\hbar\sum_{j\tau}C_{i\sigma,j\tau}\frac{dZ_{j\tau}}{dt} = \frac{\partial\mathcal{H}}{Z^{*}_{i\sigma}}.
\end{equation}
The matrix $C_{i\sigma,j\tau}$ ($i,j=1,2,\dots,A$ and $\sigma,\tau=x,y,z$) is a Hermitian matrix defined by
\begin{equation}\label{Eq5}
C_{i\sigma,j\tau} = \frac{\partial^2}{\partial Z^{*}_{i\sigma}\partial Z_{j\tau}}\log\langle\Phi(Z)|\Phi(Z)\rangle,
\end{equation}
and $\mathcal{H}$ is the expectation value of the Hamiltonian after the subtraction of the spurious kinetic energy of the zero-point oscillation of the center of mass of fragments \cite{ono1992}
%~\cite{ono1992,ono1992-1},
\begin{equation}\label{Eq6}
\mathcal{H}(Z) = \frac{\langle\Phi(Z)|H|\Phi(Z)\rangle}{\langle\Phi(Z)|\Phi(Z)\rangle}-\frac{3\hbar^2\nu}{2M}A+T_0[A-N_F(Z)],
\end{equation}
where $N_F(Z)$ is the fragment number, $T_0$ is $3\hbar^2\nu/2M$ in principle but treated as a free parameter for an overall adjustment of the binding energies.
The Hamiltonian in AMD is given in a non-relativistic form as 
\begin{equation}\label{Eq7}
H = \sum^A_{i=1}\frac{{\bf p}^2_i}{2M}+\sum_{i<j}\upsilon_{ij},
\end{equation}
where $M$ is the nucleon mass and $\upsilon_{ij}$ is the potential energy between particle \it{i} and particle \it{j}. In the present application, the standard Gogny force~\cite{Decharg1980}
%[J. Dechargé and D. Gogny, Phys. Rev. C 21, 1568 (1980)]} 
is used as the effective interaction.
%}

%\textcolor{blue}{
%The wave packet parameter $Z$ do not have physical meaning when wave packets overlap with each other such as inside a nucleus because of the effect of the antisymmetrization. Therefore physical coordinates W=$\{{\bf W}_i;i=1,\dots,A\}$ are defined approximately as
%\begin{equation}
%\mathbf{W}_i = \sqrt{\nu}{\bf R}_i+\frac{i}{2\hbar\sqrt{\nu}}{\bf P}_i = \sum^A_{j=1}(\sqrt{Q})_{kj}{\bf Z}_j,
%\end{equation}
%with
%\begin{equation}
%Q_{kj} = \frac{\partial \ln\langle\Phi(Z)|\Phi(Z)\rangle}{\partial({\bf Z}^*_k\cdot{\bf Z}^*_j)}.
%\end{equation}
%}

%\textcolor{blue}{
The NN collision process is treated as a stochastic process using the above physical coordinates at each time step. The NN collision rate is determined by a given NN cross section under Pauli principle. The NN cross section is given by
\begin{equation}\label{Eq8}
\sigma(E,\rho) = \min \left(\sigma_{LM}(E,\rho),\frac{100~mb}{1+E/(200~MeV)}\right),
\end{equation}
where $\sigma_{LM}(E,\rho)$ is the cross section given by Li and Machleidt~\cite{Li1993,Li1994}.
%~\cite{Li1993[G. Q. Li and R. Machleidt, Phys. Rev. C 48, 1702 (1993)],Li1994[G. Q. Li and R. Machleidt, Phys. Rev. C 49, 566 (1994)]}. 
The angular distribution of proton-neutron scattering are parameterized as
\begin{equation}\label{Eq9}
%\begin{split}
\frac{d\sigma_{pn}}{d\Omega} \propto 10^{-\alpha(\pi/2-|\theta-\pi/2|)}, \\
%\alpha = \frac{2}{\pi}\max \{0.333\ln E[\text{MeV}]-1, 0\},
%\end{split}
\end{equation}
\begin{equation}\label{Eq10}
%\begin{split}
%\frac{d\sigma_{pn}}{d\Omega} \propto 10^{-\alpha(\pi/2-|\theta-\pi/2|)}, \\
\alpha = \frac{2}{\pi}\max \{0.333\ln E[MeV]-1, 0\},
%\end{split}
\end{equation}
while the proton-proton and neutron-neutron scatterings are assumed to be isotropic.
%}

%\textcolor{blue}{
%The dynamical effect of the quantum fluctuations in the Gaussian wave packet is treated
%in the diffusion (and shrinking) process in the time evolution of the nucleon propagation~\cite{Ono1999,Ono2002}.
%As described in details in the references, this process is taken into account in order to treat properly the %multifragmentation process.
%In the present simulations, the version in Ref.~\cite{Ono1999} is used and it is called AMD/D in this article.
%}

%\textcolor{blue}{
For cluster production studies, a version of AMD is made in which cluster formation is treated as the final states of the two body collision process. The version is called AMD-Cluster described in details in Ref.~\cite{ono2019}.
%~\cite{ono2019 [Prog. Part. Nucl. Phys. 105, 139-179 (2019]}
An application of this version for the experimental data for the $^{12}$C+$^{12}$C reaction at 50 MeV/nucleons was made and the results were presented in one of our previous publications~\cite{Han2020}.
%~\cite{Han2020 [Han et al. PHYSICAL REVIEW C 102, 064617 (2020)}
%}

\subsection*{II-2. Fermi boost and 3NC process}
%\textcolor{blue} {
The stochastic collision process described in the previous subsection is performed, using the centroid of the Gaussian wave packet. Thus the Fermi motion is only taken into account as an average energy and no explicit Fermi motion is taken into account in the collision process.
%}
In our previous works as presented in Refs.~\cite{lin2016,wada2017}, the inclusion of the Fermi-momentum fluctuation in the collision processes and the 3NC process were studied for the energetic proton production at the incident energies from 44 to 137 MeV/nucleon, based on the modified versions of the antisymmetrized molecular dynamics (AMD) model described in the previous subsection.
%In AMD, a given reaction system is expressed by a Slater determinant of N Gaussian wave packets, where N is the nucleon number of the system. Because of the antisymmetrization, AMD can make nuclei at their ground state. However, since the Fermi motion is taken into account only as an average energy to reproduce properly their experimental binding energy, the momentum of each nucleon in a given initial nucleus is minimal, which leads to very stable ground state nuclei, but causes very low productions for nucleons on the high energy side. 

In order to take into account the Fermi motion explicitly in the dynamical evolution, two stochastic processes are incorporated in the original AMD formulation, which introduces fluctuations in the reaction. One is for the fluctuation during the time evolution of the wave packets in a given effective interaction~\cite{ono1996} and the other is the momentum fluctuation added in the collision process~\cite{lin2016}. These two stochastic processes made significant improvements to reproduce the experimental data up to the incident energies below 50 MeV/nucleon as presented in Refs.~\cite{ono1996,ono1999,lin2016}. 

The former process is an extension of the study of the particle emission from the excited nucleus. Since the equations of motion solved using the centroids of the Gaussian wave packets, the particles emission from an excited nucleus reveals the classical nature, that is, the emission probability increases linearly as the excitation energy increases. When a momentum fluctuation is added in each time step, which is evaluated from the distribution of the Gaussian wave packet, interpreting the distribution as the probability distribution of the momentum of each nucleon, the quantum nature of the particle emission is restored~\cite{ono53.845.1996}. The quantum diffusion process is an extension of the momentum fluctuation, which is added in phase space according to the Vlasov equation to minimize the deviation of the energy conservation. This modified AMD code is made by A. Ono {\it et al.}, and called AMD/D~\cite{ono1996}. 

The above treatment is mainly related to the time evolution of the wave packet diffusion process and significantly affects on the multi-fragmentation process, but contributes little for the high energy nucleon emissions. An additional momentum fluctuation, therefore, is added in the collision process to reproduce the high energy component of the experimental proton energy spectra at the incident energy up to 50 MeV/nucleon in Ref.~\cite{lin2016}. The process is called Fermi Boost and this version of AMD is called AMD/D-FM. 
In the actual calculation for given coordinate vectors $\textit{\textbf{r}}_1$ and $\textit{\textbf{r}}_2$ of two attempt colliding nucleons, the associated momenta $\textit{\textbf{P}}_1$ and 
$\textit{\textbf{P}}_2$ are given similarly to Ref.~\cite{lin2016} as
\begin{equation}\label{Eq11}
\textit{\textbf{P}}_{i}=\textit{\textbf{P}}_{i}^{0}+\Delta \textit{\textbf{P}}'_{i}\ \ \ (i=1,2).
\end{equation}
$\textit{\textbf{P}}_{i}^{0}$ is the centroid of the Gaussian momentum distribution for the particle $i$. The second term $\Delta \textit{\textbf{P}}'_{i}$ is the Fermi momentum randomly given along the Gaussian distribution.
Since the momentum distribution is partially taken into account in the wave packet propagation through the diffusion process, the following form of $\Delta \textit{\textbf{P}}'_{i} $ is taken.
\begin{eqnarray}\label{Eq12}
\nonumber \Delta \textit{\textbf{P}}'_{i} &=& \frac{|\Delta \textit{\textbf{P}}_{i}| - c P_{PF}}{|\Delta \textit{\textbf{P}}_{i}|} \Delta \textit{\textbf{P}}_{i}\ :\ 
(|\Delta \textit{\textbf{P}}_{i}| > c P_{PF}),\\ 
%&=& 0.\ \ \ \ \ \ \ \ \ \ \ \ \ \ \ \ \ \ \ \ \ \ \ :\ \ (|\Delta \textit{\textbf{P}}_{i}| \leq c P_{PF}).  \number
&=& 0.\ \ \ \ \ \ \ \ \ \ \ \ \ \ \  :\ \ (|\Delta \textit{\textbf{P}}_{i}| \leq c P_{PF}).  
\end{eqnarray}
$c P_{PF}$ is a correction term to avoid a double counting with the diffusion process. 
%\textcolor{blue}{
At the incident energy above 290 MeV/nucleon, the choice of the c factor is not very sensitive and c=0.3 is used in the present calculation. 
%}
$P_{PF}$ is a Fermi momentum and $P_{PF}=250$ MeV/c is taken.

\begin{eqnarray}\label{Eq13}
\Delta \textit{P}_{i\tau} &=& \hbar\sqrt{\nu}(\rho_{i}/\rho_{0})^{1/3}G(1).
\end{eqnarray}
$G(1)$ is a random number generated along the Gaussian distribution with $\sigma$ = 1.  $(\rho_{i}/\rho_{0})^{1/3}$ in Eq.(\ref{Eq13}) is used for taking into account the density dependence empirically. $\rho_i$ is the density at $\textit{\textbf{r}}_i$ and $\rho_{0}$ is the normal nuclear density. The index $\tau$ corresponding to the $x, y, z$ coordinates.

\begin{figure}[ht]
    \centering
    \includegraphics[width=6cm]{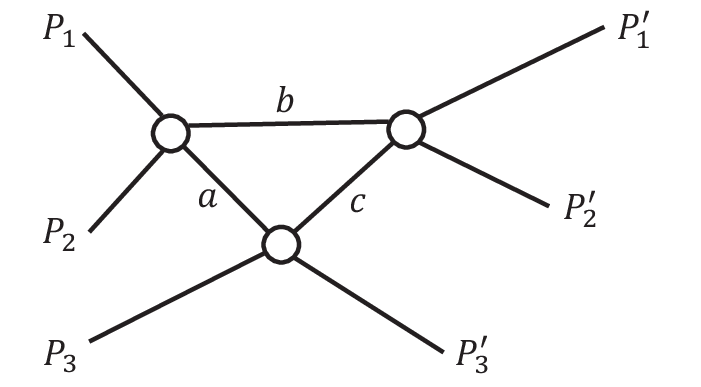}
    \caption{Diagrammatic representation of 3N collision. The lines
    indicate particle trajectories and the meeting points indicate the
    location of the particles at the time of collision. $\bf{P}_{1},\bf{P}_{2},\bf{P}_{3}$ represent the
    initial states and $\bf{P}_{1}^{'}, \bf{P}_{2}^{'}, \bf{P}_{3}^{'}$, 
    are the final states. $\bf{a, b, c}$ are intermediate
    states treated as virtual states in the 3N collision process.
    }
    \label{fig:3N_diagram}
\end{figure}

When these models were applied to the experimental data around the incident energies of 100 MeV/nucleon, however, significant deviations were observed between the available experimental proton energy spectra and those of the simulations, as presented in Ref.~\cite{wada2017}. In order to reproduce the experimental data, a 3NC process is incorporated in AMD/D-FM, following the work of Bonasera {\it et al.} in the extended BNV~\cite{bonasera1994}. The Fermi Boost is added for the three nucleons when all three pairs of nucleons are within a collision distance, which is evaluated with a constant nucleon-nucleon (NN) cross section of 40 mb. The 3NC process is performed in a similar manner to the NN collisions in AMD,
following the Bonasera's formulation in Ref.~\cite{bonasera1991} and the 3NC process is performed along the diagram shown in Fig.~\ref{fig:3N_diagram}~\cite{bonasera1994}. 

In the diagram the 3NC process is described by a succession of three binary collisions when three nucleons are in the collision distance with each other at the initial stage. Pauli principle is respected only at the final states, $\bf{P}_{1}^{'}$, $\bf{P}_{2}^{'}$, $\bf{P}_{3}^{'}$, but not at the intermediate states, $\bf{a, b, c}$. The 3NC probability increases as the density increases, pointed out in Ref.~\cite{mrowczynski1985} and therefore it occurs mostly at the early stage of the reaction before the generated hot-high density nuclear matter expands. As demonstrated in Refs.~\cite{germain1997,wada2017}, the 3NC term enhances the high energy nucleon production. This is simply because in 3NC, the kinetic energy of the three nucleons can be shared among them. This makes a significant difference in the high energy proton energy spectra between the simulations with turning ON/OFF for the 3NC process, which is demonstrated in Ref.~\cite{wada2017}. Therefore the high energy nucleon emission mechanism in the 3NC term is purely kinematic effect in the code. This modified version of AMD is called AMD$/$D$-$3NC.

%To demonstrate the sensitivities for the constant c in Eq.(\ref{Eq12}) and the density dependence in Eq.(\ref{Eq13}) at the incident energies studies in this work, the simulated results are compared with different values of the c factor and ON/OFF of the density dependence. AMD simulations of $^{12}$C+$^{12}$C at 400 MeV/nucleon with AMD$/$D$-$3NC are used. Details of the comparisons between the experimental data and simulations will be discussed in Sec.III, but one can see here that the dependence of the c factor is small, especially at larger angles, and the density dependence is only effective at the forward two angles to obtain slightly better results. 

%\end{figure}
%\textcolor{blue}{
These two processes become effective at different incident energy regions. Fermi boost becomes important for the high energy nucleon production in the heavy ion reactions below 50 MeV/nucleon, and 3NC process becomes effective in those around 100 MeV/nucleon and above. For cluster productions and their studies, AMD/D works very well. Note that all of the AMD versions described in this article are far from perfect yet and each has good and poor parts. Therefore a proper version should be applied according to the reaction studies and the reaction systems at a given incident energy.
%}

\subsection*{II-3. Semi-relativistic AMD}
\begin{figure} [ht]
    \centering
    \includegraphics[width=\linewidth]{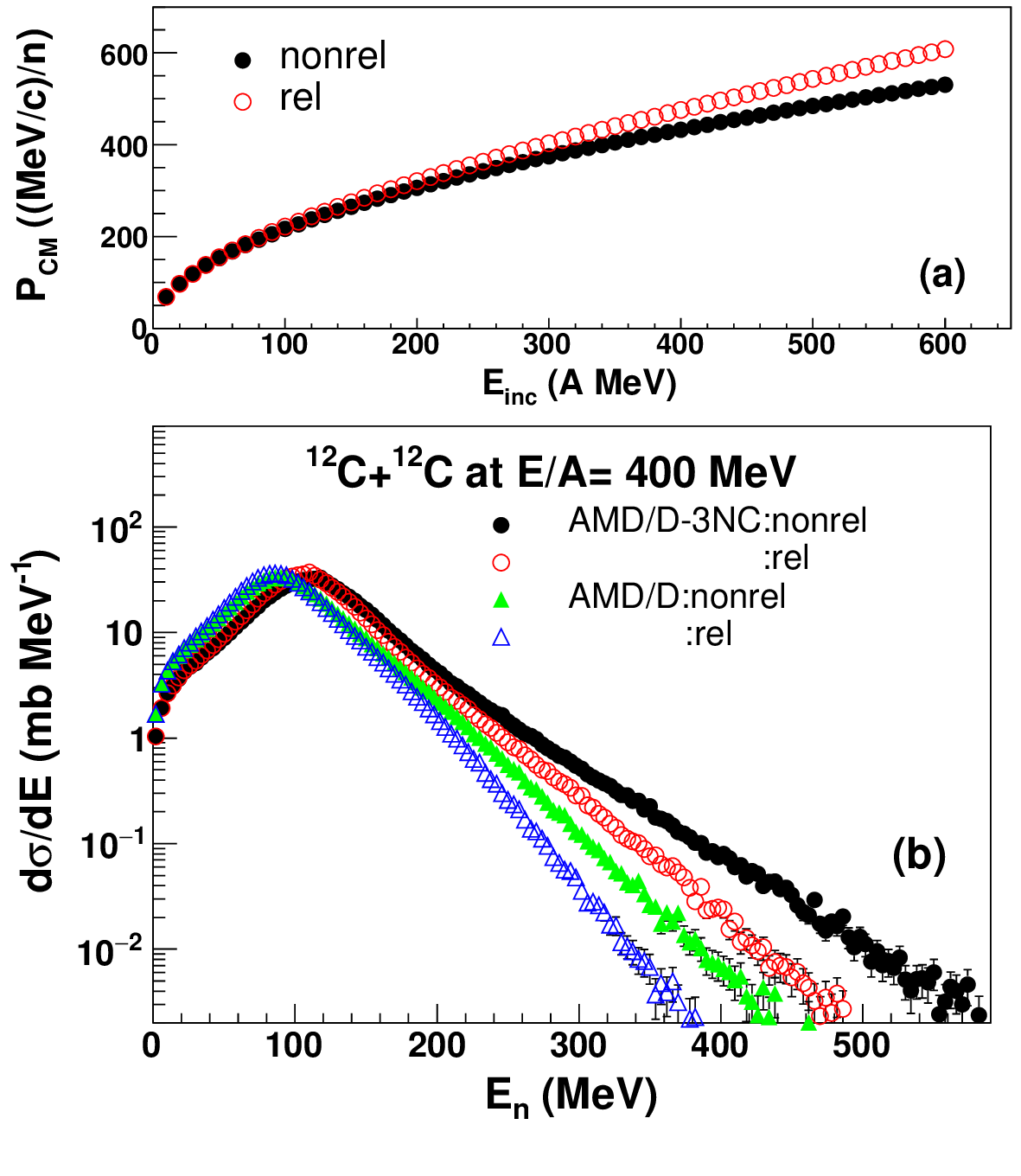}
    \caption{(Color online) Non-relativistic and relativistic energy calculations are compared for (a) the CM momentum and (b) neutron kinetic energy in the CM frame for $^{12}$C+$^{12}$C at 400 MeV/nucleon. 
    }
    \label{fig:fig00}
\end{figure}

At the incident energies, $E_{inc}/A  \geq 100$ MeV, the relativistic effect becomes non-negligible. 
%\textcolor{blue}{
In the original AMD 
%in Ref.~\cite{ono1992} 
simulations are performed, using the 
non-relativistic Hamiltonian, given in Eq.(\ref{Eq7}). In the present work we keep this non-relativistic formulation, following Ref.~\cite{Ikeno2019}.
%~\cite{Ikeno2019 Ikeno_AMD_JAM_PRC93p044612_2016}. 
There are two crucial parameters whose values may cause noticeably changes in the neutron energy spectra in the laboratory frame between non-relativistic and relativistic treatments. One is the center of mass (CM) momentum and the other is the neutron kinetic energy calculation.
%The velocity of the center of mass (CM) frame becomes more than a half of the light velocity at $E_{inc}/A \geq 500$ MeV in the non-relativistic calculation, as shown in Fig.~\ref{fig:fig00} (a) by black closed circles. At $E_{inc}/A = 600$ MeV, the velocity is reduced about 30\% in the relativistic calculation. On the contraryto the CM velocity 
The center of mass (CM) momentum increases as shown in Fig.~\ref{fig:fig00}(a) for the relativistic calculation.
%} 
After the non-relativistic simulations are performed according to the formulation presented in Section II-1, the neutron kinetic energies are calculated in the relativistic and non-relativistic forms and presented in Fig.~\ref{fig:fig00}(b) for AMD/D and AMD/D-3NC. In each case the relativistic calculation of the neutron kinetic energy is reduced about 20\% at $E_{kin} \ge 200$ MeV compared to those from the non-relativistic calculation.

For AMD/D, the high energy neutron yields exponentially fall off more rapidly above 200 MeV. This indicates that the high energy neutron productions are dominated by the incorporated stochastic processes, especially by the 3NC process, and they are independent of the non-relativistic or relativistic formulation except for the total energy conservation after the process is performed. The total energy restoration is performed by making slight shifts in phase space of those among the surrounding nucleons and rather insensitive to the high energy neutron production. Therefore in the following scenario for the relativistic treatment, the non-relativistic form is kept in the AMD simulations and the relativistic corrections are performed for these simulated results. 

\begin{figure} [ht]
    \centering
    \includegraphics[width=\linewidth]{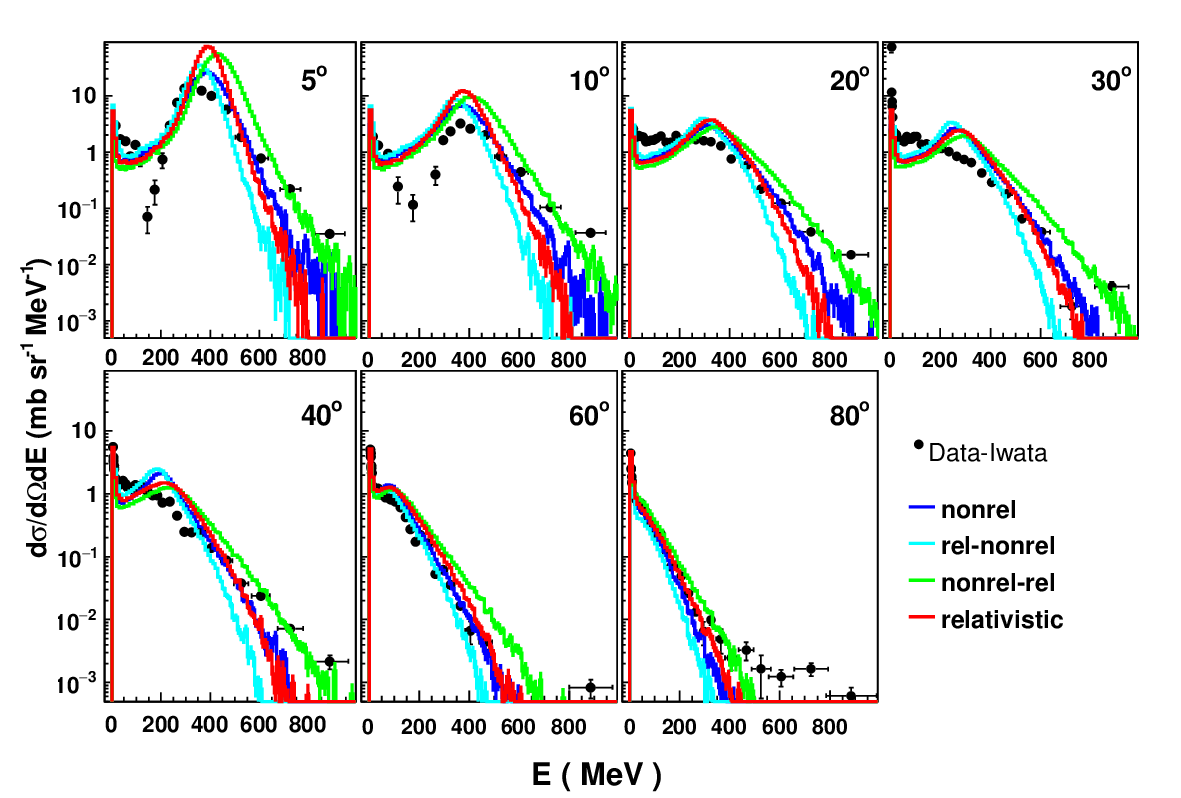}
    \caption{(Color online) The three steps of the relativistic treatments are presented separately for AMD/D-3NC simulations for $^{12}$C + $^{12}$C at 400 MeV/nucleon. Blue, light blue, green and red histograms represent the results of non-relativistic, the front end alone (denoted as rel-nonrel), the back end alone (nonrel-rel) and all three steps corrections (relativistic), respectively. The data are taken from Iwata's data set in ~\cite{iwata2001,iwata2001-1} 
    }
    \label{fig:fig000}
\end{figure}

In the original AMD in Ref.~\cite{ono1992} simulations are performed non-relativistically in the CM frame. In order to compare the neutron energy spectra and angular distributions with the experimental data, the simulated results have to be transformed into the laboratory (LAB) frame. The relativistic corrections are made in three steps. At the beginning of the calculation (front end), the input system is boosted from the LAB frame to the CM frame. At the end of the simulation (back end), the same boost, but with opposite sign, is applied. Since the CM momentum boost becomes larger for the relativistic treatment as shown in  Fig.~\ref{fig:fig00} (a),  
the relativistic transformation at the front end reduces the neutron energy and that at the back end enhances the neutron energy. 
At the back end, the relativistic form is also used to calculate the neutron energy as presented in Fig.~\ref{fig:fig00} (b), which results in significant reduction of the neutron energy. Each correction makes significant effects on the neutron energy spectra.

To demonstrate the effect at each correction, the changes of the energy spectra are presented in Fig.~\ref{fig:fig000} with the relativistic correction at the front end alone, with the two corrections at the back end alone and with all three corrections separately, using AMD/D-3NC calculations for $^{12}$C+$^{12}$C at 400 MeV/nucleon. The original non-relativistic neutron energy spectra are shown by blue histograms. Those with corrections at the front end alone and at the back end alone are shown by light blue and green histograms, respectively. At forward angles, large deviations from the non-relativistic spectra are made in both corrections and the deviations are slightly larger at the back end. At larger angles both deviations become smaller, especially at the front end.

Since the corrections tend to compensate in each other, the final energy spectra both with front and back end corrections become similar to those of the non-relativistic ones as presented by the red histograms. There are some noticeable differences observed at forward angles, but they are rather marginal. 
In the following analysis, these three step corrections are made for the non-relativistic AMD calculations and the simulated results are denoted as semi-relativistic AMD (Sr-AMD). The relativistic treatment taken in this work is semi-relativistic, since the corrections are made for the kinematics of the non-relativistic simulated events, and nucleon-nucleon inelastic and meson production processes are not taken into account.

%\textcolor{blue}{
A similar application of the non-relativistic AMD is made in Ref.~\cite{Ikeno2019} for the theoretical study of the nuclear symmetry energy for the pion production in $^{132}$Sn+$^{124}$Sn at 300 MeV/nucleon. In their study, AMD is combined with JAM, a relativistic transport model, in which the former treats the time evolution of the wave packets in the mean field and the latter is used for the meson production, especially for pions. The kinematic connection between these two models are made within the relativistic formulation, but the non-relativistic Hamiltonian in Eq.(\ref{Eq7}) is used in the AMD simulation part~\cite{Ikeno2023}.
%~\cite{Ikeno2023 [private communication]}
%}

\section*{III. Simulations and Results}

The semi-relativistic AMD simulations are applied to reproduce the available experimental data to study the production mechanisms of the high energy neutrons.

In experiments, precise measurements of neutron production double differential cross sections are a big challenge, especially for those with energy above 100 MeV/nucleon. Among the available experimental data sets as mentioned in the introduction,  
we utilized the data sets of $^{12}$C + $^{12}$C at 290 and 400 MeV/nucleon and $^{16}$O + $^{12}$C at 290 MeV/nucleon~\cite{iwata2001,satoh2011} in this study. The data set in Ref.~\cite{iwata2001} is referred as Iwata's data and those in Ref.~\cite{satoh2011} as Satoh's data below. 
Neutrons with energies up to about 900 MeV, were measured by the time-of-flight method. In both experiments, the experimental setups were very similar to each other. 
The experimental data are combined if they are available in both experiments. Their measured angles were slightly different for the reaction systems between $\theta_{lab} = 5^{\circ}$ and $90^{\circ}$. 

AMD/D-FM and AMD/D-3NC are applied for these reactions with a standard Gogny effective interaction and a constant 3NC cross section of 40 mb, which corresponds to the hard core nucleon-nucleon scattering and is same as that in Refs.~\cite{bonasera1991,bonasera1994}. In Ref.~\cite{bonasera1991}, the three body cross section N3 is given as
\begin{equation}\label{Eq14}
 N3=\frac{16}{3\pi}\sigma^{5/2}\rho^{3}\sqrt{T/m}V, 
\end{equation}
where $\sigma$ is the nucleon-nucleon cross section, $\rho$ is the density, $T$ is the temperature, $m$ is the nucleon mass and $V$ is the system volume. 
%\textcolor{blue}{
One should note that N3 in Eq.(\ref{Eq14}) is evaluated for a uniform nuclear matter. In the AMD simulations, when three nucleons meet within a collision distance, the surrounding density and temperature together with the front factor are dynamically simulated in the time evolution of the wave packets. Therefore the actual 3N collision cross section, $\sigma_{3N}$, used in the code is simply given by $\sigma_{3N}=\sigma^{5/2}$.
%}
About one to two million events are simulated for each case in the impact parameter range of b=0-8 fm. For $b > 8$ fm very few collisions are observed. The semi-relativistic corrections are made for all AMD/D-FM and AMD/D-3NC simulated results and they are denoted as Sr-AMD-FM and Sr-AMD-3NC, respectively.

\begin{figure} [ht]
\centering
\includegraphics[width=\linewidth]{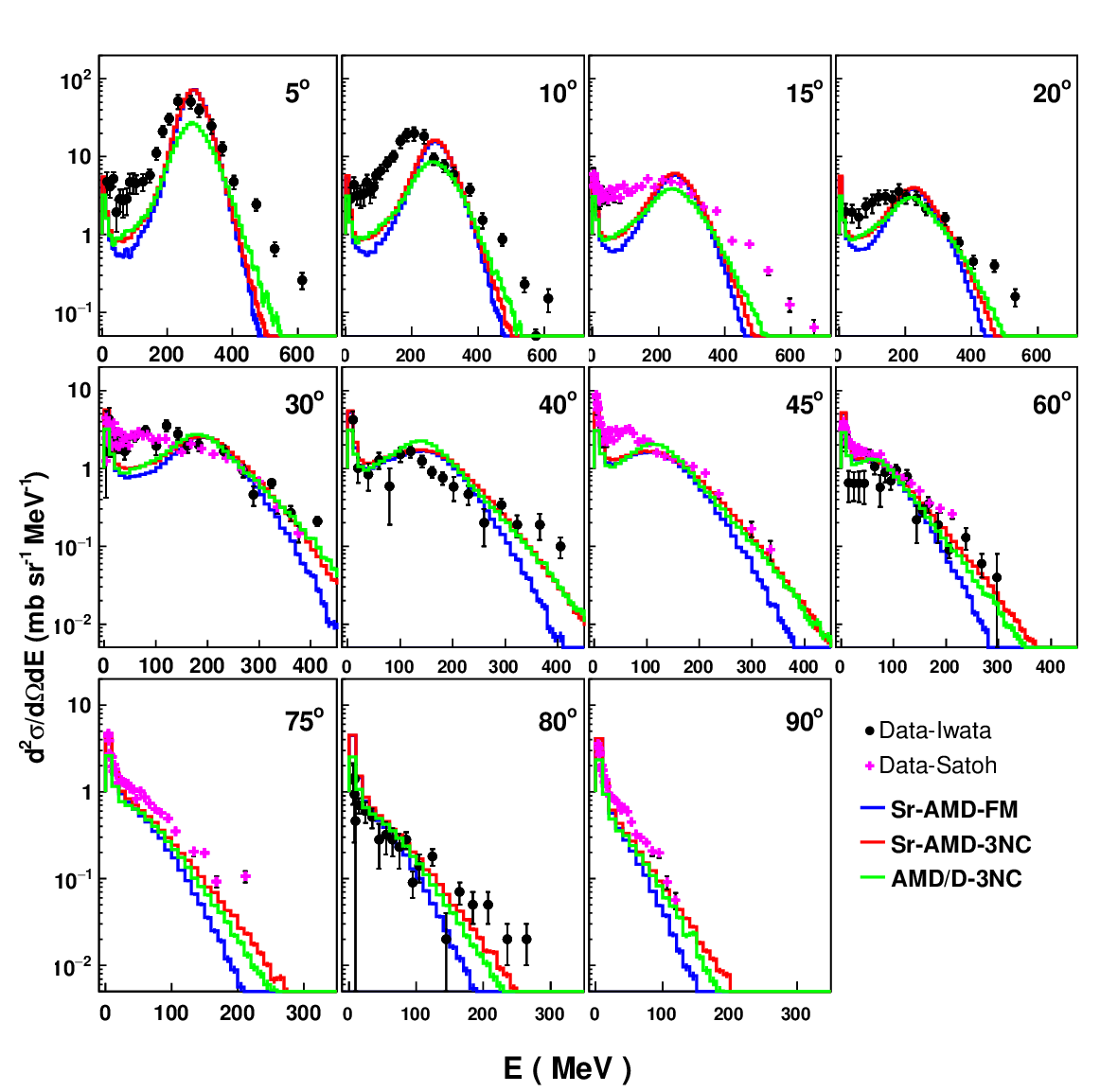}
\caption{(Color online) Neutron double differential cross sections for $^{12}$C+$^{12}$C collisions at 290 MeV/nucleon. Black solid circles and purple crosses are the experimental data taken from Refs.~\cite{iwata2001,iwata2001-1} and Ref.~\cite{satoh2011}, respectively.
Blue, red and green histograms are calculated with Sr-AMD-FM, Sr-AMD-3NC and AMD/D-3NC, respectively.
}
\label{fig:c12c12_290AMeV}
\end{figure}

In Fig.~\ref{fig:c12c12_290AMeV}, the simulated and experimental results for the $^{12}$C+$^{12}$C reaction at 290 MeV/nucleon are compared over the observed angles. All results are plotted in an absolute scale. The high energy neutrons at larger angles are reasonably well reproduced both by AMD/D-3NC and Sr-AMD-3NC, especially for those from Iwata's data, whereas Sr-AMD-FM predicts much soft high energy tails at these angles. 
The Satoh's data well agree with those of Iwata's data at 30$^{\circ}$ where both measurements were made at the same angle, whereas at 75$^{\circ}$ and 90$^{\circ}$, the Satoh's data show about twice larger cross sections than those of the calculations. On the other hand, the double differential cross sections at 80$^{\circ}$ from Iwata's data are well reproduced by both AMD/D-FM and AMD/D-3NC at low energies and indeed these cross sections are smaller by a factor of about two, compared to those of Satoh's at 75$^{\circ}$ and 90$^{\circ}$. Therefore the origin of the discrepancies between the experimental data and the simulations at 75$^{\circ}$ and 90$^{\circ}$ are inconclusive for its origins either from the experiments or from the AMD simulations. The neutron spectra from Iwata's data at 40$^{\circ}$, 60$^{\circ}$ and 80$^{\circ}$ are reasonably well reproduced with the 3NC process in the entire energy range.
At $5^{\circ}$ the peak yield is better reproduced by Sr-AMD, but the experimental peak energy is slightly low and its width is slightly wider. The experimental yields on the low energy side at $\theta \leq 30^{\circ}$ are significantly larger than those of all simulations.
\begin{figure} [ht]
    \centering
    \includegraphics[width=\linewidth]{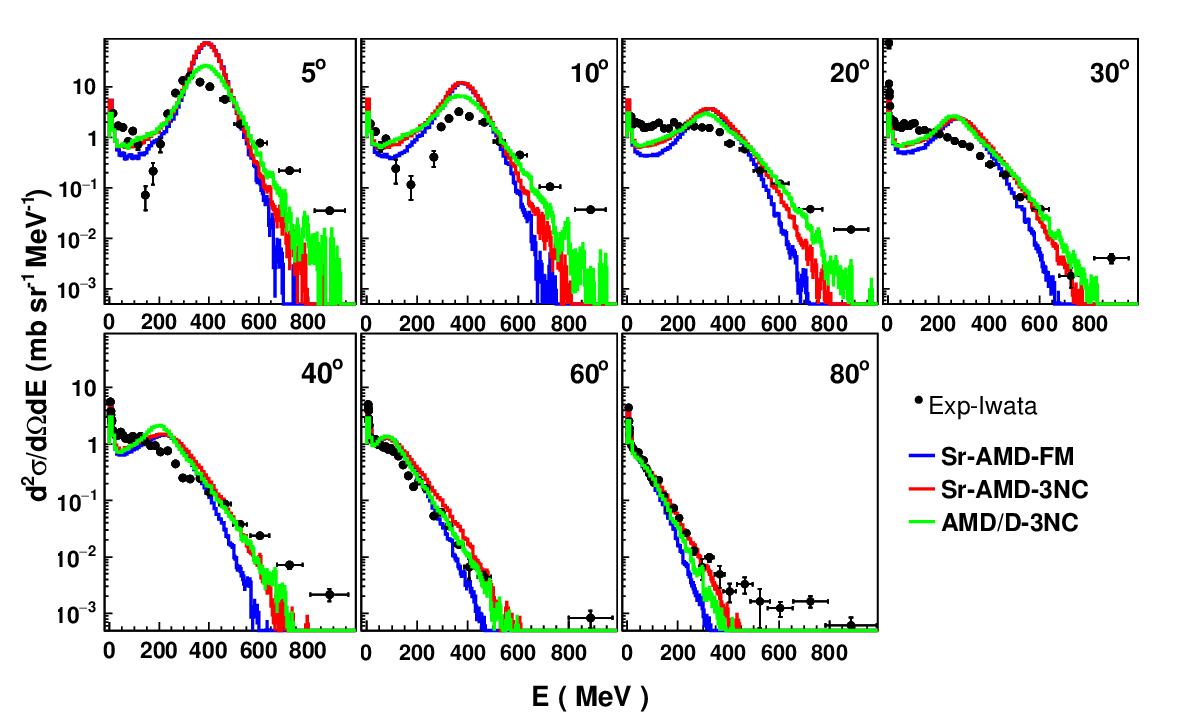}
    \caption{(Color online) Neutron double differential cross sections for $^{12}$C+$^{12}$C collisions at 400 MeV/nucleon. Black solid circle points are experimental data taken from Ref.~\cite{iwata2001}. 
    Histograms are same as those in Fig.~\ref{fig:c12c12_290AMeV}.
    }
    \label{fig:c12c12_400AMeV}
\end{figure}

In Fig.~\ref{fig:c12c12_400AMeV}, the simulated neutron spectra are plotted with the experimental $^{12}$C+$^{12}$C reaction at 400 MeV/nucleon from $\theta_{lab} = 5^{\circ}$ to $80^{\circ}$. 
The high energy neutrons at $\theta \geq 60^{\circ}$ are rather well reproduced both with AMD/D-3NC and Sr-AMD-3NC, whereas Sr-AMD-FM predicts slightly softer high energy tails at these angles. On the low energy side, significant overpredictions are observed. At $\theta =5^{\circ}$ and $10^{\circ}$ all AMD simulations overpredict the yields but at $20^{\circ} \leq \theta \leq 40^{\circ}$ they underpredict the yields.  At angles $\theta \leq 40^{\circ}$, all AMD simulations show a pronounced quasi-elastic peak with about twice larger cross sections. This feature is quite contrast to the experimental results, especially at $20^{\circ} \leq \theta \leq 40^{\circ}$. The experimental data do not show any peak structure but show a rather broad shoulder. This may cause the significant underpredictions on the lower energy side. On the contrary, at $\theta = 5^{\circ}$ and $10^{\circ}$, the experimental data show peaks with similar widths, but 2-3 times less yields.  These discrepancy patterns are quite different from those in Fig.~\ref{fig:c12c12_290AMeV}. The different discrepancy patterns in peak position and amplitude between the simulations and the experimental results at these forward angles may suggest that they are caused by the experiments. 

\begin{figure} [ht]
    \centering
    \includegraphics[width=\linewidth]{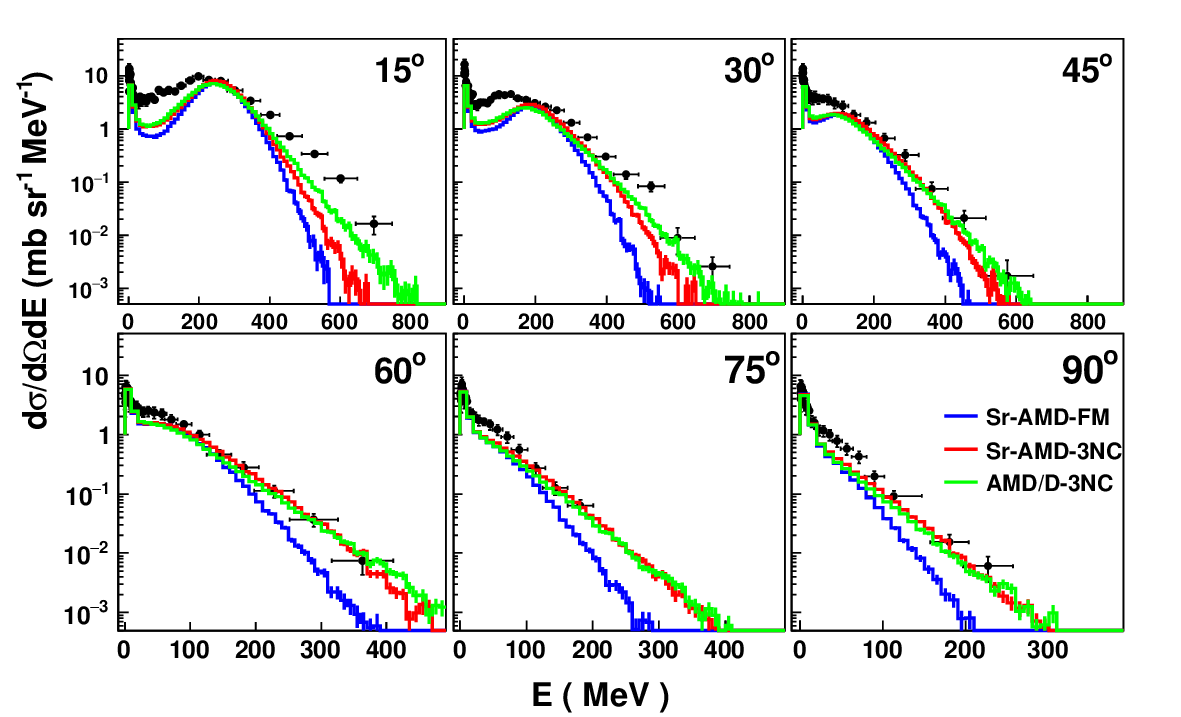}
    \caption{(Color online) Neutron double differential cross sections of $^{16}$O+$^{12}$C collisions at 290 MeV/nucleon. Black solid circles are the experimental data taken from Ref.~\cite{satoh2011}.Blue, red and green histograms are same as those in Fig.~\ref{fig:c12c12_290AMeV}.}
    \label{fig:16o12c_290AMeV}
\end{figure}

The semi-relativistic and non-relativistic AMD models are also applied to $^{16}$O+$^{12}$C at 290 MeV/nucleon and the results are shown in Fig.~\ref{fig:16o12c_290AMeV}. The experimental high energy neutron tails are well reproduced in overall with the 3NC process in this case, except at $\theta = 15^{\circ}$. At $\theta = 15^{\circ}$ the experimental tail is harder than the non-relativistic one. Sr-AMD-FM again predicts significantly softer tails than the experimental ones. At angles $\theta \leq 45^{\circ}$, the experimental low energy yields are significantly underpredicted by about a factor of two for all three simulations.

For the above three reaction systems, non-relativistic AMD (green histograms) and Sr-AMD (red) both with 3NC are resembles in each other, especially at larger angles. A missing relativistic treatment in Sr-AMD is the time evolution of the wave packet in the effective mean field. As discussed in Sec.II-3, it is expected that the relativistic time evolution will not change the neutron spectra so much, since this does not affect on low energy neutrons and high energy neutrons are mostly generated by the incorporated stochastic processes. Therefore the above results indicate that the non-relativistic AMD is still valid in the incident energy range studied here. 
%\textcolor{blue}{
However the similarity between the non-relativistic and semi-relativistic results obtained in this study is not guaranteed at higher incident energies nor in other heavier reaction systems. We actually observed noticeable effects of the semi-relativistic treatments in the data analysis of the incident energies 560-600 MeV/nucleon at large angles, but these results will be presented in our future works. 
%}

\section*{IV. Summary}

AMD/D-FM and AMD/D-3NC are applied for the high energy neutron productions in light heavy ion collisions, using the $^{12}$C+$^{12}$C and $^{16}$O+$^{12}$C reactions at 290 and 400 MeV/nucleon. The relativistic corrections are made to the non-relativistic AMD simulations to apply them to the experimental data at these incident energies. The semi-relativistic version of AMD/D-3NC and AMD/D-FM are applied to the above experimental data as well as the original non-relativistic ones. The final semi-relativistic results end up similar to the original non-relativistic ones. For all cases, the high energy neutron tails are well reproduced by AMDs with 3NC. AMD/D-FM simulations significantly underpredict these high energy neutron productions.  These observations, therefore, suggest that the high energy neutrons with energy above the incident beam energy per nucleon are mainly produced by the 3NC process. In overall these results are consistent to the results obtained around $E_{inc}/A \sim 100$ MeV. On the other hand, all simulations fail to reproduce the low energy neutrons below the beam energy at angles around $20-45^{\circ}$, though the reason for the discrepancies below 15$^{\circ}$ are left inconclusive either from the experiments or from the simulations. 
In order to confirm our results of the 3NC process in the energetic nucleon emissions, further precise experiments are necessary in the future. The High Intensity heavy ion Accelerator Facility (HIAF)~\cite{yang2013}, which is being built in China, as well as other facilities such as FRIB and RIKEN, will provide us the opportunity to probe the emission mechanism of high energy nucleons and sub-threshold particles. 

\paragraph{Acknowledgements}
We would like to thank Prof. A. Ono for providing the AMD/D source codes and Dr. Y. Iwata for providing their numerical data sets of the experimental cross sections for all reaction systems. 
%\textcolor{blue}{
We also thank Dr. N. Ikeno for her helpful comments about their work.
%} 
We acknowledge Dr. Hao Qiu for reviewing the manuscript. We also thank Dr. J. Z. Duan for the help of computing. This work is supported by the Strategic Priority Research of Chinese Academy of Sciences, Grant No. XD34030000, the National Natural Science Foundation of China under Grant No. U1832205 and the US Department of Energy under Grant No. DE-FG02-93ER40773.


\begin{thebibliography}{**}
\bibitem{iwata2001-1}Private communication. the plotted data are from the numerical data set, which were kindly sent to us from Dr. Y. Iwata.
\end{thebibliography}
\end{document}